# Single-ion Kondo behavior of Ce in a novel Kondo lattice, CeNi$_9$Si$_4$

**Kausik Sengupta and E.V. Sampathkumaran**
*Tata Institute of Fundamental Research, Homi Bhabha Road, Colaba, Mumbai – 400005, India.*

**Abstract**

The compound, CeNi$_9$Si$_4$, has been recently reported to be an unusual Kondo lattice with a large Ce-Ce separation, with a breakdown of Kadowaki-Woods relationship governing low temperature electrical resistivity and heat-capacity. Here we report the results of magnetic susceptibility, heat-capacity and electrical resistivity of the solid solution, Ce$_{1-x}$La$_x$Ni$_9$Si$_4$, to understand the Kondo behavior of this compound. The results establish that the observed properties of Ce in this compound are single-ionic in character.

In most of the Ce intermetallic compounds known till to date, the Ce-Ce distance is close to 4 Å and, generally speaking, this is the critical value of Ce-Ce separation below which one sees characteristic Kondo coherence features in the physical properties of Kondo lattices at low temperatures. It is usually difficult to find integer-valent (3+) Ce compounds in which Ce-Ce spacing is large enough so that intersite correlations can be ignored [1]. In this respect, the compound, $CeNi_9Si_4$, crystallizing in a fully ordered tetragonal (space group I4/mcm) variant of the cubic $NaZn_{13}$, gains importance, as the fraction of Ce ions is rather small (just about 7 atomic percent) and Ce-Ce separation is quite large 5.6 Å. Michor et al [2] reported the results of magnetic, thermodynamic and transport properties of this compound and claimed that this material is a non-magnetic Kondo lattice with an extraordinary position among Kondo lattices, as Ce-Ce Kondo intersite interaction can be ignored for this compound. Interestingly enough, this compound is characterized by a remarkably low ratio (about $\sim 10^{-7}$ $\mu\Omega$ cm(molK/mJ)$^2$) of $A/\gamma^2$ (where A is the coefficient of $T^2$ in electrical resistivity ($\rho$) and $\gamma$ is the linear term in heat-capacity (C)), compared to that expected on the basis of Kadowaki-Woods relation [3]. We have therefore undertaken the investigation of the La substituted alloys, $Ce_{1-x}La_xNi_9Si_4$, by bulk measurements to establish that the properties, including the above-mentioned $A/\gamma^2$ anomaly, are single-ionic in character.

It is worthwhile mentioning that the temperature dependencies of heat capacity and magnetic susceptibility ($\chi$) can be accounted [2] for theoretically only if one assumes that Ce is fully degenerate, that is j = 5/2. Integral-valent Ce intermetallics are characterized by a low Kondo temperature ($T_K$), which is generally much lower than the crystalline electric field (CEF) splitting. Therefore, the ground state doublet only is mainly involved in the Kondo effect with an effective total angular momentum j = 1/2 for T<<$T_K$. In intermediate-valence compounds, stronger hybridization between 4$f$ and conduction electrons leads to high $T_K$ ($\sim$100 K) values and, only in such Ce compounds, usually the full total angular momentum j = 5/2 needs to be considered for the Kondo effect. Thus, the compound $CeNi_9Si_4$ is an example for a trivalent Ce-based compound with j = 5/2.

The samples, $Ce_{1-x}La_xNi_9Si_4$ (x= 0.0, 0.2, 0.4, 0.6, 0.8 and 1.0), were prepared by arc melting stoichiometric amounts of constituent elements followed by homogenization in an evacuated sealed quartz tube at 950 C for 7 days. The samples were characterized by x-ray diffraction (Cu $K_\alpha$) and found to be single phase. The lattice constants obtained (from selected peaks at higher angle side) are shown in Table 1. It is clear from this table that there is a monotonic increase of lattice constants with La substitution for Ce, interestingly with a dramatic increase when x is varied from 0.4 to 0.6. The $\chi$ measurements (1.8-300 K) were performed with the help of a superconducting quantum interference device (Quantum Design, USA). The $\rho$ data were obtained by a conventional four-probe method. Temperature (T) dependent C behavior (1.8-50K) was tracked by the relaxation method with the help of a commercial set-up (Quantum design, USA).

The results of $\chi$ measurements taken in a field of 5 kOe are shown in figure 1 for all compositions. The $\chi$ is found to be essentially temperature independent typical of Pauli paramagnets for La compound (4.12 x $10^{-4}$ emu/mol). In the case of Ce containing compositions, there is a gradual increase with decreasing temperature with a tendency to flatten around 50 K, which is cut off by an upturn at lower temperatures. The observed low temperature tail for the parent Ce compound is more prominent than in Ref. 2 and, therefore, this tail is attributed to paramagnetic impurities. In order to get an idea of this impurity contribution and to subtract off this part from the measured $\chi$, we fitted the data below 50 K to the expression, $\chi(T) = \chi(0) + nc/T$, where $\chi(0)$ is the $\chi$ value as T tends to zero, and the second term represents paramagnetic impurity part. We assume that the impurity term arises from trivalent Ce ions, either from a small separate phase or from Ce ions with an imperfect crystallographic environment (due to disorder). This means that 'c' takes a value of 0.807 emu/mol and 'n' is molar fraction of (trivalent) Ce. From such a fitting, we deduce that the impurity content is not more than 0.7%. The corrected $\chi$ values obtained after subtracting this Curie term due to the impurity are shown in figure 2 after normalizing to Ce content. It is clear from this figure that there is a broad peak in $\chi(T)$ in figure 2 around 50 K for x= 0.0, in agreement with the report of Michor et al [2]. It is well-known [4] that the peak temperature is related to $T_K$ with the Kondo effect arising from the fully degenerate trivalent Ce. (There is another peak at much lower temperatures, which we believe is an artifact of a small error in the subtraction of impurity contribution). It is to be noted that the peak temperature does not vary as Ce sub-lattice is diluted with La, thereby implying that there is no profound change in the Kondo behavior of Ce among the pseudo-ternary alloys. In the data thus derived, below about 180 K, there is a gradual deviation from the high temperature Curie-Weiss region, possibly due to gradual domination of Kondo compensation effects (see below for the estimates of $T_K$); the values of paramagnetic Curie temperature ($\theta_p$), which is known to be related to $T_K$, and the effective magnetic moment were obtained (see table 1) from this linear region (180-300 K) of the inverse $\chi$ versus T plot. The values of $\theta_p$ are found to fall in the range -75 and -100 K. The magnitude of $\theta_p$ increases marginally from -75 K for x= 0.0 to -100 K for x= 0.8, and this is unexpected considering that the expansion of the lattice caused by La substitution should have suppressed the value of $T_K$ (as well as indirect exchange interaction strength, $T_{RKKY}$). Possibly, the $\chi$ data well beyond 300 K may have to be measured to look for Curie-Weiss region in a wider temperature range, as the value of $T_K$ could actually be large compared to that of $\theta_p$. A qualitative comparison of the variation of the strength of the Kondo effect can be also obtained by normalizing the $\chi(Ce)$ values to Ce concentration and it is obvious that the curves in figure 2 nearly overlap for x = 0.0, 0.2 and 0.4, whereas the ones for x= 0.6 and 0.8 deviate slightly upwards from the rest, as though the value of $T_K$ for these compositions are lower by about 5 to 10% compared to Ce rich compositions. A better idea about the trends in the x-dependence of $T_K$ can be inferred more precisely from the $\chi(1.8K)$ values from the analysis proposed by Rajan [5]. The values of $T_K$ thus determined (see table 1) is almost constant (falling in the range 180 – 190 K) for x = 0.0 to 0.6 within the limits of experimental error (about 5%); however, for x = 0.8, there is a fall to about 140 K, as though lattice expansion at the La-rich end caused a marginal depression of $T_K$. From this analysis, one can claim that $T_K$ is almost independent of x at least till x = 0.6. The large values of $T_K$ thus inferred from this analysis may endorse a point made above that the Curie-Weiss region employed is not reasonable to infer trends in $T_K$ from $\theta_p$.

For the sake of completeness, we have also measured isothermal magnetization (M) at 1.8 and 50 K for all compositions, the results of which are shown in figure 3. While the plots are linear at 50 K, there is a small curvature at low fields (below 20 kOe) at 1.8 K with a linear variation at high fields. The value of M even at high fields is rather small. Considering this, it can be concluded that the magnetic impurity is negligibly small to influence our conclusions.

We now discuss heat capacity behavior shown in figure 4. As expected, there is no evidence for magnetic ordering in the temperature range of measurement. A fitting of the data for La compound in the range 1.8 to 10 K to the formula, $C = \gamma T + \beta T^3$, yields a value of about 32 mJ/mol-K$^2$ and about 400 K for $\gamma$ and Debye temperature respectively, in good agreement with Michor et al [2]. This value of $\gamma$, unusually large for La compound, appears to indicate that this compound by itself is a heavy-fermion possibly arising from a finite 4$f$ occupancy and the resultant correlation with the conduction band. It is not clear whether this heavy-fermion behavior results from the correlations within the 3d band of Ni. In the case of the parent Ce compound, a similar fitting yields a value of about 180 mJ/mol-K$^2$, which is marginally higher than that reported in the literature (155 mJ/mol K$^2$). The present results confirm that this compound is a heavy-fermion. Now, turning to the influence of La-substitution on the heavy-fermion behavior, it is found that $\gamma$ per mol decreases monotonically with increasing dilution of Ce sublattice (see table 1); however, while $\gamma$ per Ce-mol falls in the range of 180 to 200 mJ/mol K$^2$ (essentially constant within the limits of experimental error) for x = 0.0, 0.2 and 0.4, the corresponding values get noticeably smaller for higher values of x. The values of $T_K$ have been determined from the relation given by Rajan [5] and shown in table 1. The values thus obtained compare quite well with those derived from the $\chi$ data. Wilson ratio $R_w$ obtained from the measured $\chi(0)$ (as T tends to zero Kelvin) and $\gamma$ values is found to be close to unity in the entire series.

In order to understand the transport anomaly of the parent Ce compound, the $\rho$ data are shown for the La substituted alloys in figure 5. The $\rho$ of the parent compound exhibits $T^2$-dependence at low temperatures (below 30 K) typical of Fermi-liquids, and the overall shape of the plot is the same as that reported by Michor et al [2]. In the case of non-magnetic Kondo lattices, the plot of $\rho(T)$ is usually characterized by the double-peaked structure if the crystal-field effects are operative, and the absence of this feature is consistent with full degeneracy of Ce ion. Clearly, reasonably large values of $T_K$ wash out crystal-field splitting. This is found to be true even when the lattice is expanded by La substitution for Ce. A question that arises is whether the gradual fall of $\rho$ below 100 K for x= 0.0 arises from the coherent scattering among Kondo centers. If this is the case, then the gradual replacement of Ce by La should destroy this fall, resulting in an upturn of $\rho$ at low temperatures, typical of Kondo-lattice to Kondo impurity transformation, as demonstrated for many Ce and Yb compounds [see, for instance, Refs 6, 7], in sharp contrast to the observation. Therefore we conclude that the shape of $\rho(T)$ in the entire T-range is single-ionic in character. It may be recalled [8] that intermediate valent Ce compounds are in fact characterized by similar S-shaped $\rho(T)$ curves and therefore this compound is an ideal example for a Ce compound at the border of trivalent Kondo lattices and intermediate valent compounds. The temperature range over which Fermi-liquid behavior is valid increases gradually from 30 K for x= 0.0 to 100 K for x= 0.8. The coefficient of $T^2$ term for each composition is given in the table, along with the ratio of A/$\gamma^2$. The points to be noted are: (i) this ratio is nearly the same as that claimed by Michor et al [2] for the parent Ce compound, thereby confirming their finding of a breakdown of Kadowaki-Woods relation, and (ii) this breakdown persists even for the substituted alloys, which establishes that this anomaly is single-ionic in character. Finally, the shape of ρ(T) curve for x= 1.0 is in agreement with that reported in Ref. 2, in which it was argued that there are significant electron-electron or paramagnon scattering effects even in the case of La compound; this inference gains credence from large γ value as well. For this reason, we are unable to derive 4f-contribution to ρ in Ce containing alloys.

To conclude, the results of magnetic susceptibility, heat capacity and electrical resistivity measurements are presented for the pseudo-ternary solid solution, $Ce_{1-x}La_xNi_9Si_4$. The results reveal that the properties of the parent compound are essentially single-ionic in character, thus establishing that this is an unusual trivalent Ce-based Kondo lattice (with j = 5/2) with very little Kondo coherence effect. In particular, there is a breakdown of Kadowaki-Woods relation with respect to $A/\gamma^2$ ratio even in dilute alloys, thereby revealing that this anomaly is also single-ionic in nature. At this juncture, it is worth stating that similar breakdown of Kadowaki-Woods relation has been shown for many Yb compounds, though there are not many examples of this kind for Ce, barring $CeSn_3$ [9] and the present system. As argued by Tsujii et al also, this anomaly could be intimately connected with the large degeneracy. We hope this will serve as a clue for the theoretical understanding of this puzzling issue.

We thank Kartik K Iyer for his help during measurements.

**Table 1:** Lattice parameters, paramagnetic Curie temperature ($\theta_p$) and the effective moment ($\mu_{eff}$) determined from the high temperature (180-300 K) Curie-Weiss region, linear coefficient of heat capacity (γ), the coefficient (A) of $T^2$ term in ρ, Kondo temperature (determined from χ(0) and γ values), Wilson ratio ($R_W$), and $A/\gamma^2$ ratio for the alloys $Ce_{1-x}La_xNi_9Si_4$.

| x | a (Å) | c (Å) | V (Å$^3$) | $\theta_p$ (K) | $\mu_{eff}$ ($\mu_B$/Ce) | $T_K^\chi$ (K) | γ (mJ mol$^{-1}$ K$^{-2}$) | $T_K^\gamma$ (K) | $R_W$ | A (x 10$^{-5}$) (mΩ cm K$^{-2}$) | $A/\gamma^2$ μΩ cm (mole K/mJ)$^2$ |
|---|---|---|---|---|---|---|---|---|---|---|---|
| 0 | 7.837 | 11.442 | 702.79 | -75 | 2.45 | 184 | 179 | 190 | 1.24 | 1.79 | 5.58E-7 |
| 0.2 | 7.839 | 11.444 | 703.24 | -84 | 2.51 | 181 | 156 | 181 | 1.19 | 0.701 | 2.88E-7 |
| 0.4 | 7.842 | 11.446 | 703.97 | -90 | 2.51 | 195 | 116 | 200 | 1.23 | 0.346 | 2.57E-7 |
| 0.6 | 7.851 | 11.461 | 706.36 | -99 | 2.50 | 181 | 111 | 142 | 0.94 | 0.427 | 3.467E-7 |
| 0.8 | 7.857 | 11.462 | 707.48 | -76 | 2.55 | 140 | 76 | 127 | 1.09 | 0.213 | 3.68E-7 |
| 1 | 7.858 | 11.462 | 707.78 | --- | --- | --- | 32 | --- | --- | 0.209 | 2.04E-6 |

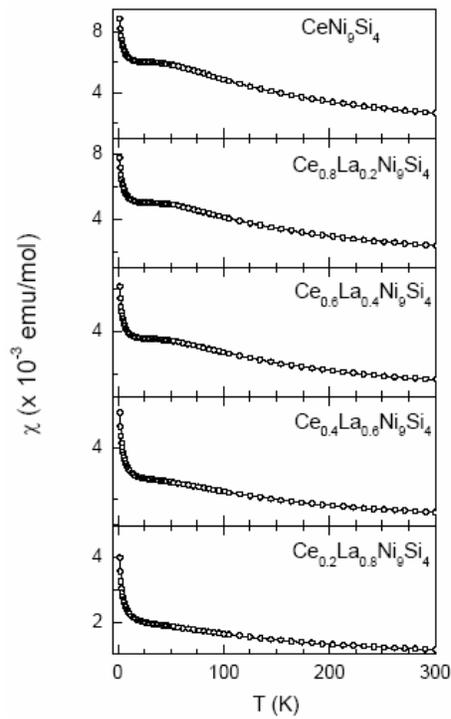

**Figure 1:** *Magnetic susceptibility as a function of temperature for the alloys $Ce_{1-x}La_xNi_9Si_4$ measured in the presence of a magnetic field of 5 kOe.*

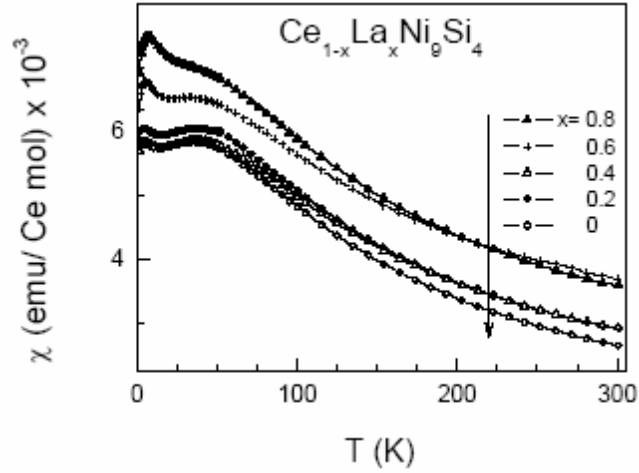

**Figure 2:** M*agnetic susceptibility (per Ce mol) after subtraction of low temperature tail for $Ce_{1-x}La_xNi_9Si_4$. The lines though the data points serve as a guide to eyes.*

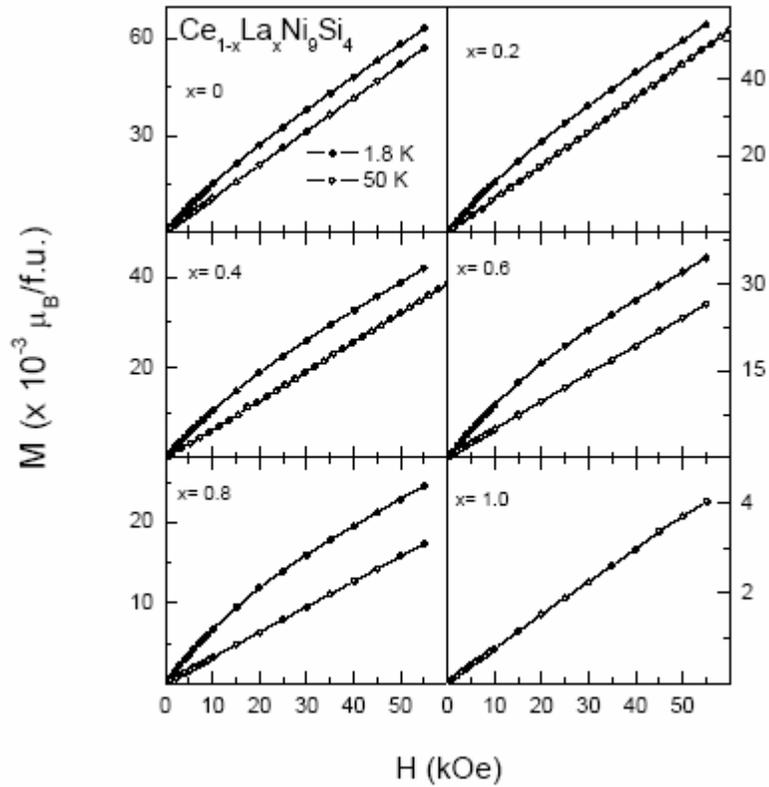

**Figure 3:** *Isothermal magnetization behavior of the alloys, $Ce_{1-x}La_xNi_9Si_4$, at 1.8 and 50 K. The lines through the data points serve as guides to the eyes.*

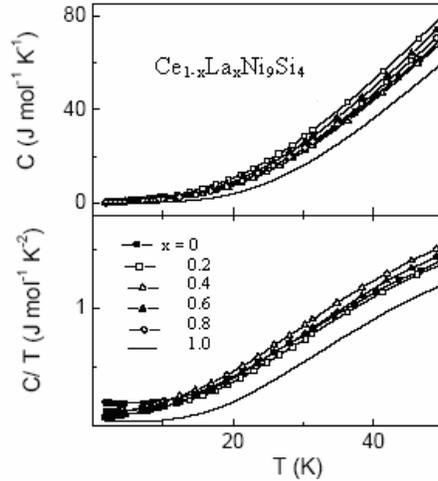

**Figure 4:** *Heat capacity data of the alloys, $Ce_{1-x}La_xNi_9Si_4$, plotted in two different ways. The lines through the data points serve as guides to the eyes.*

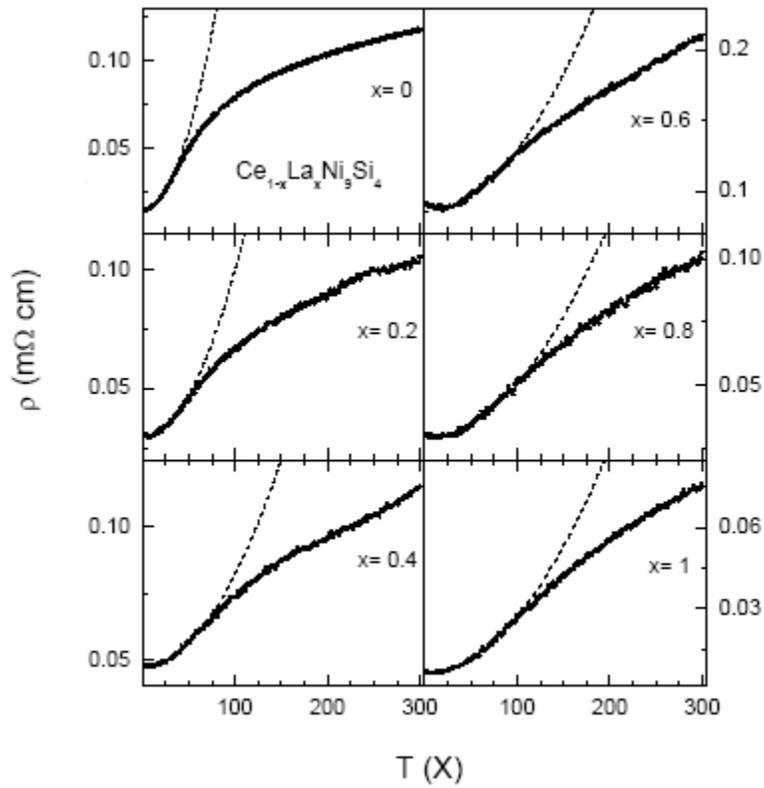

**Figure 5:** Temperature dependence of electrical *resistivity for $Ce_{1-x}La_xNi_9Si_4$. The dashed lines are fit of the equation $\rho = \rho_o + AT^2$.*